MAY 2023

# Putting collective intelligence to the enforcement of the Digital Services Act

Report on possible collaborations between the European Commission and civil society organisations

## Dr Suzanne Vergnolle

Associate Professor in Technology Law
Conservatoire National des Arts et Métiers

# Contents

## List of abbreviations

| | |
|---|---|
| **Art.** | Article |
| **CSOs** | Civil Society Organisations |
| **DSA** | Digital Services Act |
| **DSCs** | Digital Services Coordinators |
| **EC** | European Commission |
| **EU** | European Union |
| **GDPR** | General Data Protection Regulation |
| **OECD** | Organisation for Economic Co-operation and Development |
| **TEU** | Treaty on European Union |
| **TFEU** | Treaty on the Functioning of the European Union |
| **VLOPs** | Very Large Online Platforms |
| **VLOSEs** | Very Large Online Search Engines |

## List of tables

## List of annexes

# Executive Summary

## Background

In 2022, the European Union adopted the DSA (Digital Services Act), a landmark regulation which aims to foster a safer digital space. The DSA sets rules for a "safe, predictable and trusted online environment".[1] More specifically, the regulation reconciles the liability exemption of intermediary services with new due diligence obligations for mitigating the risks intermediary services create for society, including phenomena like hate speech, discrimination, and disinformation.[2]

Some of the new rules have real potential to improve online services practices, but their actual impact will only be as good as their implementation and enforcement.[3] Enforcement involves various actors in a maze of roles and responsibilities. Each Member State has to designate one or more authorities tasked with enforcing the DSA which will then be coordinated by the national Digital Services Coordinator. National supervisory authorities are responsible for overseeing the providers of intermediary services, including online platforms established in their countries.[4] For services with more than 45 million monthly active users in the European Union,[5] designated as VLOPs (Very Large Online Platforms) or VLOSEs (Very Large Online Search Engines), the Commission (European Commission) has exclusive supervision of their due diligence obligations.

As national and EU authorities get ready to take on complex tasks to enforce the regulation, CSOs (Civil Society Organisations) are advocating for strong enforcement, and getting organised to assist regulators in this endeavour.

## Objectives

The aim of this report is to provide organisations, policy makers, and regulators with key recommendations and expert advice on cooperation mechanisms in order to ensure CSOs are fully involved in the DSA's enforcement. By its nature, the enforcement system relies on multiple institutions both at the national and European level and their combined work will be essential to ensure effective enforcement.

Because Member States can adopt different models of enforcement, ranging from one DSC to multiple authorities coordinated by the DSC, it is difficult to provide recommendations that can apply across all Member States. This report consequently focuses primarily on formulating recommendations for CSOs to support the Commission's work. Hoverer, many of these recommendations are easily transposable to national authorities.

Most importantly, since the European Commission is gearing up to use its new enforcement powers,[6] it is both timely and relevant to formulate recommendations on how CSOs can support this important work and actions.

## Key findings

The Digital Services Act is an important piece of legislation likely to contribute to the protection of several important rights and interests. Contrary to previous legislation in this domain, its enforcement is complex because it involves multiple regulatory authorities (competent national authorities, Digital Services Coordinators, Board, European Commission). The main regulatory authority for the enforcement against VLOPs and VLOSEs is the Commission, which has to develop investigation and enforcement capabilities from the ground up.

Various implicit references in the DSA encourage cooperation with civil society organisations. One of the key references is article 64 of the DSA which requires the Commission, in cooperation with the Digital Services Coordinators and the Board, to develop Union expertise and capabilities. While this article does not impose a specific way to do so, it opens up the possibility of establishing an expert group.

Establishing an expert group specifically dedicated to support the effective application and enforcement of the DSA is not just a possibility, it is something that is widely recommended. Also, this creation is aligned with the growing trend inside authorities to involve stakeholders at every stage of legislation and to rely on external expertise for monitoring, implementation, and enforcement. This is also fully in line with the Commission's wish to build a culture of consultation and dialogue.

---

6 Thierry Breton, the European Commissioner, instructed the "Commission teams to enforce DSA no later than: 1 September 2023": see his tweet of 19 December 2022.

The Commission's decision of 2016 establishing a general framework for the creation and operation of expert groups proves to be a good structure, with important principles such as transparency by default, fair representation, and distribution of powers. Because an expert group focusing on the application and enforcement of legislation has some particularities, the composition of the group should be carefully considered, and only certain types of experts should be allowed to apply. More broadly, thoughtful design of the group's operation and logistics are essential to ensure involvement and the success of this group.

Aside from their participation in the expert group, civil society organisations can contribute to the enforcement of the DSA in other ways. One of the most efficient ways for them to do so is by notifying regulators of potential violations with formal complaints. Another way is to analyse documents and data produced under the transparency obligations. Indeed, as a "data-generating piece of legislation", the DSA provides a great opportunity for civil society organisations and individuals to support the analysis of the available data.

## Key recommendations

While underlying the many ways to build strong cooperation settings between regulators and CSOs, this report focuses on making concrete recommendations for the design of an efficient and influential expert group with the European Commission. The creation of an expert group finds its roots in article 64 and recital 137 of the DSA which require the Commission to develop Union expertise and capabilities. Once established, the experts of this group will be able to bring evidence-based information directly to the Commission and specific expertise on the protection of fundamental rights and the safety of users online. By instituting an expert group, the Commission will not only benefit from valuable expert knowledge but will also demonstrate its willingness to put in place an efficient enforcement system based on collective intelligence.

Aside from the establishment of an expert group, other cumulative mechanisms will also help the DSA's enforcement to thrive. Civil society organisations should, for instance, consider organising regular crowdsourcing events to deep-dive and analyse the data published by entities covered by the transparency obligations. As it has done in the past, the Commission can sponsor these events and be a direct beneficiary of their results. Another way for civil society organisations to bring information to the Regulator is by legal action, including by making complaints to the regulators.

A detailed list of recommendations can be found throughout this report, but a selection of ten recommendations follows:

**1** The European Commission should establish an expert group to support the effective application and enforcement of the Digital Services Act

**2** The expert group should be of a reasonable size (around 30) and consist of independent experts, civil society organisations, the Board, and the European Commission

**3** The expert group's mandate should be broadly defined. Its specific missions and priorities should be co-decided by the members of the group

**4** The expert group's Secretariat and Chairpersonship should involve representatives of the Commission and representatives of civil society organisations

**5** The secretariat should involve experts in the drafting process of the agenda and documentation

**6** The expert group should hold regular meetings (in person and remote)

**7** The Commission should offer the experts working in the group the opportunity to ask for compensation

**8** The Commission can fund the group with the supervisory fees paid by VLOPs and VLOSEs

**9** Civil society organisations should organise regular crowdsourcing events to offer analysis of the documentation made available under the DSA

**10** Regulators should put in place efficient mechanisms to receive information and process complaints

# Report

# I

# Introduction

**1. Overview.** This study is concerned with finding possible models for civil society involvement in the enforcement of the DSA. An analysis of the regulation shows numerous provisions in which potential synergies can be grounded.[7] Based on research and interviews with experts in the fields, this report identifies the need and basis for establishing an expert group. It also provides the key logistical and organisational elements needed to involve civil society organisations in the enforcement of the DSA. It particularly focuses on the enforcement at the European level because Member States will implement the DSA differently in their national laws, making recommendations general, and therefore less relevant and also because the Commission's new role is crucial in the effective enforcement of the Regulation against VLOPs and VLOSEs. However, though the recommendations are specific to the European Commission, Member States are also encouraged to draw from the conclusions of this report.

**2. Definition of enforcement.** Generally, the notion of enforcement refers to the "process of making people obey a law or rule".[8] In this report, the notion of enforcement will be taken broadly. Enforcement will refer to all the activities relating to chapter 4 of the DSA, namely activities such as:

- **monitoring**, i.e. ensuring platforms are compliant with the DSA,

- **investigating**, i.e. assessing and examining a suspected infringement,

- **deciding**, i.e. when the authority closes its investigation (notably with a non-compliance decision),

- **implementing the decision**, i.e. making sure the infringement is not perpetuated (including *via* periodic penalty payment).

---

7    See Annex I: Overview of the DSA's enforcement system.

8    Cambridge Dictionary, Online Dictionary. For a definition of enforcement in its legal sense, see OECD, "Regulatory enforcement and inspections", *OECD Best Practice Principles for Regulatory Policy*, 2014, p. 11.

**3. Definition of civil society organisations**. Civil society organisations are the principal structures of society outside of government and public administration and industry.[9] The term "civil society organisations" is inclusive and is rooted in the democratic traditions of the Member States. There is no common or legal definition of the term which is often broadly considered as "a range of organisations which include: the labour-market players (…); organisations representing social and economic players, (…); non-governmental organisations (NGOs), which bring people together in a common cause, such as environmental organisations, human rights organisations, charitable organisations, educational and training organisations, etc.; community-based organisations (CBOs) (…)".[10] Throughout the report, the notion of civil society organisation will be captured broadly to include structures independent from the state and the industry.

**4. DSA's provisions on enforcement.** The DSA is a complex law with numerous provisions dedicated to its enforcement. Annex I of this report provides a comprehensive presentation of the DSA's enforcement system.[11] The Annex also offers a critical analysis of the DSA's enforcement provisions where CSO involvement is either mandated or possible.

**5. The DSA's provisions on cooperation with civil society organisations.** The DSA includes multiple explicit and implicit references encouraging cooperation between regulatory authorities and civil society organisations. Enforcement authorities will have to develop cooperation mechanisms at the compliance level, as well as during the monitoring phase and the enforcement stage. Annex I of this report provides a list of cooperation measures with civil society organisations wich is summed up in the following charts.

---

**TABLE I:** Explicit references to the involvement of CSOs at various stages of the DSA's implementation

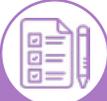

| 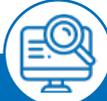  **Compliance stage** | 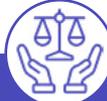  **Monitoring stage** | 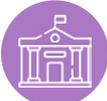  **Enforcement stage** |
|---|---|---|
| CSOs can contribute to:<br>● Risk assessments<br>● Drawing up codes of conduct<br>● Drawing up crisis protocols<br>● Developing Commission expertise<br><br>CSOs can be designated as:<br>● Trusted flaggers | CSOs can contribute to:<br>● Conducting scientific research | CSOs can be designated as:<br>● Representatives of service recipients |

**TABLE II:** Implicit references to CSO's involvement at the enforcement stage

| | | Authorities can cooperate with CSOs by: |
|---|---|---|
| 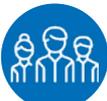  **National authorities** | **Investigation stage** | ● Sending requests for information |
| | **Enforcement stage** | ● Receiving observations |
| 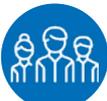  **Board** | **Monitoring stage** | ● Inviting experts to attend its meetings<br>● Cooperating with CSOs in its tasks<br>● Developing and implementing standards, guidelines, reports, templates and codes of conduct<br>● Relying on CSOs when identifying emerging issues |
| 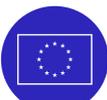  **Commission** | **Investigation stage** | ● Sending request for information<br>● Doing interviews and taking statements<br>● Designating CSOs as experts for inspection<br>● Designating CSOs as independent external experts |

**6. The resources of the European Commission.** The supervision and enforcement roles attributed to the Commission are noteworthy. As the executive power of the European Union, it is not common for the Commission to be given enforcement powers.[12] To tackle this new challenge, the Commission is strengthening its teams within the DG Connect and developing its new European Centre for Algorithmic Transparency. To do so, the Commission has been shifting present staffing resources and is expected to ramp up recruitment. In total, the Commission will hire approximately 50 new professionals from all backgrounds (legal, technological, economics, etc).[13] This number seems inadequate to properly implement and enforce the DSA, especially the aspects linked to fundamental rights protections. Therefore, while it is essential to have in house teams working on enforcement, cooperation with outside expertise will be necessary to effectively enforce the DSA.

**7. Methodology used for the elaboration of this report.** The various recommendations presented were developed following a four-step method.

First, broad research on how to involve stakeholders into the work of regulatory authorities was conducted, showing an extensive variety of mechanisms.

Second, specific research on different models for advisory bodies was conducted. In the EU context, advisory bodies can take multiple names such as expert groups, multi stakeholders, or an observatory. They are based on some recurring principles: they are established by a public body to "provide ongoing advice on matters requiring substantive scientific and technical analysis, and whose membership consists largely of experts drawn from non-government organisations and research centres".[14] Advisory groups exist inside and outside public administration, always with the goal of offering assistance and providing strategic advice.[15] Setting up an advisory body is a traditional, yet very valuable, way for institutions to build on expertise from stakeholders.[16]

---

12   In the Competition law field, the Commission has already overseen anti-competition practices, mergers, and state aid.

13   To be precise, Thierry Breton announced the need to "staff the dedicated DG CONNECT team with over 100 full time staff, combining both DSA and DMA". See Thierry Breton, "Sneak peek: how the Commission will enforce the DSA & DMA – Blog of Commissioner Thierry Breton", EU Commission website, 5 July 2022.

14   For a general and historic presentation on expert advisory bodies, see Kate Crowley and Brian Head, "Expert Advisory Bodies in The Policy System", in *Routledge Handbook of Comparative Policy Analysis*, 2017, p. 186.

15   According to a 2014 BDC study, 86 percent of entrepreneurs who have an advisory board say "it's had a significant impact on their business", BDC, "How an advisory board can boost your business", bdc website.

16   For a general and historic presentation on expert advisory bodies, see Kate Crowley and Brian Head, "Expert Advisory Bodies in The Policy System", in *Routledge Handbook of Comparative Policy Analysis*, 2017, p. 181 et s.

Third, interviews were conducted. After having identified various advisory bodies set up by the European Commission where civil society organisations are involved, such as the Multistakeholder expert group to support the application of the GDPR,[17] the Consumer Policy Advisory Group,[18] or the Advisory Committee on equal opportunities for women and men,[19] as well as other advisory bodies, such as the EU Observatory on the Online Platform Economy,[20] the author of this report conducted interviews with some of their members (active and former) to grasp the good practices and difficulties experts have encountered. The author also interviewed experts in the fileds of participation and democracy as well as those focusing on digital policies. Finally, the author interviewed public officials involved in the enforcement of rules in digital services.[21] The interviews usually lasted an hour. Sometimes additional follow-up discussions occurred via email. The interviews typically started with a general presentation of the goals of the report, which then opened up into a general discussion with the interviewees. Four topics (legitimacy,[22] administration,[23] impact,[24] and role[25]) were discussed, with detailed questions. Finally, the discussion often covered other types of CSO involvement such as complaints, fellowship programmes, and crowdsourcing events.

Fourth, based on the research and interviews, the author sought common ground on which to base this report and put forward the following recommendations.

---

17  See the **webpage** of the group.
18  See the **webpage** of the group.
19  See the **webpage** of the group.
20  See the **webpage** of the group.
21  The full list is available in **Annex II: List of interviews**.
22  The main idea was to make sure the group is representative of multiple voices.
23  The main idea was to make the participation's group as easy and valuable as possible.
24  The main idea was to make the group impactful for all parties involved.
25  The main idea was to discuss the technicity of the meetings.

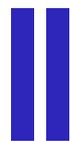

# Overview of the existing participatory models

**8. Overview.** This section will briefly explain why transparency and participation are important in a democratic society (A), and how democratic values play out at the global and EU level. It will then present some of the existing cooperation mechanisms between CSOs and institutions (B).

## A. Preliminary remarks

**9. Importance of transparency and participation.** That governments should be transparent, participatory, and accountable to the public is not a novel idea. Member States have recognised, long before the creation of the European Union, the importance of transparency and accountability for governments. For example, France declared in its famous Declaration of Human and Civic Rights, "Society has the right to ask a public official for an accounting of his administration",[26] opening the path for government accountability.

**10. At the global level.** Transparency, accountability, and openness in government actions are increasingly recognised as central to economic development and political stability.[27] At the international level, the Open Government Partnership, the multilateral initiative committed to making governments more open, accountable, and responsive to citizens, is a simple, yet strategic way for civil society and reformers in governments to join forces in monitoring commitments taken by governments.[28] Open government is a cornerstone of an open society, a society where voices are heard, ideas debated, and where there is opportunity for exchange between the government and the people.[29]

---

International law recognizes the right to participation in policymaking as a human right under article 25 of the International Covenant on Civil and Political Rights.[30]

**11. At the European level.** As the European Union developed from an organisation primarily based on economic values to one that combines economic, social, and political objectives, its policy has increasingly turned to value based issues such as democracy, governance, and human rights.[31] Article 2 of the TEU (Treaty on European Union) explicitly refers to the founding values of the European Union which are (among others) human dignity, freedom, democracy, and the rule of law. Under the Treaties, institutions should be open, transparent and cooperate with external stakeholders. Article 15 of the TFEU (Treaty on the Functioning of the European Union) recognises civil society's role in the EU's good governance[32] and article 11 of the TEU stresses the need for the institutions to have an open, transparent and regular dialogue with civil society organisations.[33] Therefore, multiple models of cooperation have been put in place in various sectors and the Commission has recognised cooperation with stakeholders as a crucial element of its policy making.[34] In other words, the Commission has been building a "culture of consultation and dialogue" with civil society organisations.[35] In parallel, the Commission has acknowledged that, although it has "considerable in-house expertise, it needs specialist advice from outside experts as a basis for sound policymaking".[36] This is particularly true in fast moving sectors where expertise very often lies outside public institutions. Recently, the Commission has published a communication on the enforcement of EU law, where it recognised the crucial role of civil society and individuals to "draw attention to possible breaches and the need for them to be addressed"[37] as well as the need for the Commission to be more transparent.[38]

---

This trend can also be found in other EU institutions, including the European Parliament. For instance, the Parliament recently published a study calling for an integrated participatory and deliberative model for the EU.[39] This study analyses the "EU participatory toolbox", including complaints from citizens to the European Ombudsman, and the realities of EU citizens' participation. These important issues include low participation, fragmentation, unequal access, and limited integration in decision making.[40] While the impact of various stakeholders in the EU policy in general is broader than the scope of this report, there is a growing trend inside and outside public institutions to reflect and act to better involve stakeholders at every stage of their work. It is particularly important to reinforce this trend because it contrasts with a background of intense lobbying, with concerns including corruption, and the reality that civil society does not have the capacity and resources to match industry lobbying.

## B. Multiple collaboration mechanisms

**12. Diversity of involvements.** Collaboration between public institutions and civil society organisations is not new, but the pace of its development has been accelerating. Citizens are asking their governments to be more transparent as well as more efficient about their services, practices, procedures, laws, and more. One of the ways to do so is by offering access to public sector information[41] as well as to reinforce collaborations with stakeholders.[42] Collaboration can take various forms and each authority has its own ways to involve third parties in its work.

The common trends of collaboration include roundtables, public consultation, committees, and conferences. Recently, with new technological developments, EU institutions have experimented with new formats of collaboration. For instance, the Publications office of the EU is holding an annual open data competition "to stimulate innovation and transform the interaction between citizens and the EU administration".[43] Other tech-oriented events, such as the EUvsVirus hackathon, have also demonstrated the positive societal impact of these new forms of cooperation.[44]

**13. Public consultations.** Public consultations are one of the most widespread ways of involving civil society in the institutions' work. The Commission is mandated by the Treaty to carry out broad consultations with parties concerned.[45] The goal is to "consult as early and as widely as possible in order to include all interested parties".[46] Public consultations can be considered a good way for civil society organisations to provide well-structured feedback and contribute to the debate with information from the ground.[47] At the same time, public consultations have a limited impact, notably because they are often conducted as a box-ticking exercise, rather than a meaningful attempt to engage and because powerful interest groups often circumvent them and lobby the institutions directly.[48] In addition, the timelines and amount of effort needed to be put into answering a consultation, including being able to know that the public consultation is held, can be prohibitive.[49] Public consultations are not a cooperation mechanism *stricto sensu* since the institution "enjoys unconstrained discretion to determine (...) what to do with their findings".[50] Therefore, in many cases, the "incentives to contribute to a public consultation remain modest, as proven by the modest number of responses submitted and the limited representativeness of the interests contributing to the consultations".[51]

**14. Expert groups.** Expert groups working with the European Commission are set up either by Decision of the Commission (formal groups) or by a Commission department after being authorised (informal groups).[52] Not all advisory bodies are mandated by law and the Commission can set advisory bodies without a strong legal basis which can be "a factor of strength for the group, testifying that the process is fully voluntary and grounded on the good will of its members".[53]

As of February 2023, there are 672 active expert groups.[54] Many of them are gathering representatives from Member States to coordinate and share exchanges of views.[55] Aside from the expert groups gathering national and European public bodies, other groups bring together a broad variety of stakeholders. It is particularly the case for the groups whose missions are to assist with the "implementation of existing Union legislation, programmes and policies".[56] Such groups include, for instance, the multistakeholder expert group to support the application of the GDPR,[57] the Consumer Policy Advisory Group,[58] or the Advisory Committee on equal opportunities for women and men.[59] Traditionally, expert groups are more focused on policy making and less on enforcement *per se*,[60] probably because the Commission is rarely charged with enforcement powers.

**15. Pools of experts.** Aside from the organised expert groups, which are consultative bodies that meet regularly,[61] regulators have also created Pools of Experts[62]. These pools of experts are usually established "with a view of providing material support in the form of expertise that is useful for investigations and enforcement activities and to enhance the cooperation and solidarity between all Supervisory Authorities".[63] Generally, they are a "reserve list of subject matter experts, from which collaborators may be selected to assist the agency in carrying out the work activities".[64] In this context, experts are only expected to be invited by the authority to contribute occasionally to specific projects. Compared to expert groups, the pools of experts provide a more flexible way to receive expertise but are highly dependent on the authority's investment and attention to them.

**16. Complaints.** At the enforcement stage, other mechanisms such as formal complaints are used by civil society organisations to get the attention of regulators on specific violations.[65] For consumer organisations, complaints have been a good way to raise public awareness and to make the organisation part of the procedure.[66] "Delegated" complaints, where consumer groups or associations represent individuals, are fairly recent and institutions have been developing more direct paths to receive and treat these complaints.[67] Aside from pure litigation in court, cooperation between institutions and civil society organisations has flourished. For instance, the CPC (Consumer Protection Cooperation Network)[68] aims at tackling widespread breaches of EU consumer laws and helps coordinate investigation and enforcement.[69] Since its creation, the network has been developing more efficient ways to work with CSOs and foster direct exchange of information with parties involved.[70] In some instance, organisations were able to cooperate in the proceedings and provide up-to-date evidence to the authorities, thus helping them to reach the most appropriate and coherent decision possible. It is therefore very important to create a good dialogue between institutions and stakeholders.

**17. Recurring events.** Some authorities are also organising events with civil society organisations on an ongoing basis. For instance, the EDPS (European Data Protection Supervisor) has been organising with EDRi and research centres the "EDPS Civil Society Summit" in the context of the Privacy Camp.[71] The EDPS have also been financially supporting the organisation of the Privacy Camp. This camp is an annual conference bringing together "digital rights advocates, activists as well as academics and policymakers from all around Europe and beyond to discuss the most pressing issues facing human rights online".[72] Such recurring events contribute to a culture of long term cooperation between regulators and civil society, fostering a better understanding between the different stakeholders but also trust in the institutions.

To sum up, there are various ways in which authorities can cooperate with CSOs and some of them can be a direct source of inspiration to ensure effective enforcement of the DSA.

---

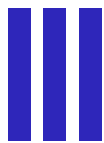

# Possible collaborations with civil society organisations for effective enforcement of the DSA

**18. Overview.** This section presents a series of recommendations which have the potential to enhance the involvement of CSOs in the enforcement of the DSA. Among the multiple models of collaboration already described, one has been considered a very efficient way to achieve a long term dialogue between authorities and CSOs: the establishment of an expert group. Unlike a pool of experts, which is a list of people who can be used for multiple purposes, the expert group is a body instituted with a specific mandate and requires specific analysis.

This section will start by discussing the opportunity of setting up an expert group, as well as recommendations on its structure and how to make it impactful and valuable (A). Then it will explore some cumulative mechanisms that could help contribute to effective enforcement of the DSA (B).

## A. Setting up an impactful and valuable expert group

**19. The DSA's references to the creation of expert groups.** The protection of fundamental rights as well as the safety of users is at the core of the DSA and indeed is listed as one of its main objectives in its first article.[73] As one of the central goals of the regulation, it is evident that the enforcement of the rules will require cooperation with organisations that have strong experience of these topics because they have been working on them for a long time. Indeed, civil society organisations can provide evidence-based recommendations to the regulators[74] and bring up the voice of people whose rights have been hurt by online intermediaries[75].

---

73   See for instance **recital 3** and **article 1** of the DSA which reads as follow: "The aim of this Regulation is to contribute to the proper functioning of the internal market for intermediary services [...] and in which fundamental rights enshrined in the Charter, including the principle of consumer protection, are effectively protected." See also, Commission, "**Digital Services Act: EU's landmark rules for online platforms enter into force**", 16 November 2022.

Also, the DSA requires the Commission to develop its own expertise and capabilities but does not mandate a specific way to do so.[76] While the DSA does not explicitly mandate the establishment of an expert group, it strongly encourages its creation. By doing so, the regulatory authorities will demonstrate their willingness to properly enforce the regulation, to draw upon the relevant expertise to do so, and to cooperate with organisations while putting in place an efficient system based on collective intelligence. For the enforcement at the EU level, the expert group can also help make sure the enforcement against VLOPs and VLOSEs is evidence based, considers the variety of interests affected by the services, and is rooted in the protection of individuals' rights and society in general.[77]

**20. Overview.** This section will discuss why it is important to establish expert groups at the national level and with the European Commission (1). It will then present key recommendations for the expert group with the European Commission, including recommendations on the composition for the group (2), its mandate (3), and its administration (4).

## 1. Establishment of expert groups at the national level and with the European Commission

**21. Overview.** Under article 64 of the DSA, "the Commission, in cooperation with the Digital Services Coordinators and the Board, shall develop Union expertise and capabilities".[78] One way to bring expertise to the Commission, as well as to the Member States, is by establishing expert groups.

**22. Two levels, one aim.** Effective enforcement of the DSA will rely heavily on national authorities because the Commission only takes on exclusive supervision on the due diligence rules in place for VLOPs and VLOSEs. While the regulatory authorities have in-house expertise, it will be very important for them to also rely on external expertise.

---

74   Julian Jaursch, "In search of strong platform oversight in the EU", *Ökologisches Wirtschaften*, 4/2022, 1 December 2022.

75   This is why article 86 of the DSA recognises the right for recipients of service to mandate civil society organisations as their representatives.

76   Art. 64 of the DSA. This obligation will be difficult to enforce because there is no actionable right to hold the Commission accountable but also because it is sufficiently vague that it will be difficult to review its implementation. Interview with Julian Jaursch, 4 November 2022.

77   On the voluntary creation of expert groups, see Federico Paoli, Ingrid Schmidt, Olivia Wigzell and Andrzej Ryś, "An EU approach to health system performance assessment: Building trust and learning from each other", *Health Policy*, 2019, vol. 123, no 4, p. 405.

78   See also recital 137 of the DSA which is very informative. For a brief analysis of this article, see below § 81.

Because CSOs have approaches, perspectives, and information that are different and complementary to those of the authorities, they can be useful, if not essential, in properly enforcing many of the DSA's provisions and contribute to achieving the DSA's aims. Therefore, expert groups should be established both within the European Commission (a) and at the Member States level (b). Interactions between the two levels should also be organised (c).

## a. Establishing an expert group with the European Commission

**23. Expert group within the Commission.** To develop "union expertise and capabilities",[79] the Commission can establish an expert group. By doing so, the Commission will be able to gather input from experts working in civil society organisations as well as other independent experts. The Commission is open to the idea of setting up an expert group to help with the enforcement of the DSA, since it is already referring to the "Expert Group on Digital Services" in public documents.[80]

## b. Establishing expert groups within Member States

**24. Expert group with the Digital Services Coordinators and other competent authorities.** By default, most of the enforcement actions will go through Member States' authorities. Under article 56 of the DSA, the "Member State in which the main establishment of the provider of intermediary services is located shall have exclusive powers to supervise and enforce this Regulation". Therefore, it is vital to swiftly and thoroughly designate entities that will "enable strong platform supervision".[81] In doing so, it will be essential for authorities to create mechanisms to involve civil society organisations at the national level in order to oversee and properly enforce the Regulation. Trusted flaggers are natural actors to consider in contributing to these bodies because of their experience and recognition in specific fields.[82] However, the group should not be limited to trusted flaggers because not all relevant civil society organisations will be designated as such.[83]

### c. Interactions between the national and European levels

**25. Multi-level cooperation on specific topics.** Even for the enforcement against VLOPs and VLOSEs, which is mainly assigned to the European Commission,[84] cooperation with Member States authorities and national civil society organisations can prove valuable. For instance, the Commission will also rely on the contributions of the DSCs for the designation of providers falling into the category of VLOPs and VLOSEs. Indeed, there is no proactive obligation for providers of online services to inform the Commission that they meet the specific threshold.[85] Providers only have to regularly publish "information on the average monthly active recipients of the service in the Union".[86] When an online platform meets the specific threshold, the DSC must inform the European Commission.[87] Therefore, the first step of the enforcement against VLOPs and VLOSEs, namely their designation, can be facilitated by national authorities. This complex designation work can easily be crowdsourced, and civil society has already created a central chart to document which platforms fall under the specific threshold.[88]

**26. Multi-level expert groups' cooperation and exchanges.** To help information and expertise flow better between the different expert groups, they should make their work available and, when necessary, organise gatherings. Communication tracks between expert groups at the European and national level can be developed, notably to resolve potential inconsistencies and ensure better cooperation. The Board should work towards making sure this cooperation is successful.

> **Recommendations**
>
> ● The European Commission should establish an expert group
>
> ● Member States should establish expert groups
>
> ● The Commission should rely on the inputs brought by the national regulatory authorities and expert groups at the national and European level
>
> ● The Board should foster a dialogue between expert groups

---

84  Under **article 56 § 2** of the DSA, the European Commission has exclusive powers to supervise and enforce the additional obligations for VLOPs and VLOSEs. However, under **article 56 § 4** of the DSA, where the Commission has not initiated proceedings, the competent national authority of the main establishment of the provider of VLOPs or VLOSEs has supervisory powers for the enforcement of the other applicable provisions.

85  The providers are obliged however to answer requests from DSCs and the Commission (**art. 24 § 3** of the DSA).

86  **Art. 24 § 2** of the DSA.

87  **Art. 24 § 4** of the DSA.

88  See below the proposals for crowdsourcing events, § 68 and s. Chart crowdsourced in a spreadsheet, see **here**.

## 2. Composition of the expert group

**27. Setting up an expert group.** Because the creation of expert groups at the Member States level depends on a variety of factors, including the institutional landscape and political differences, the subsequent discussion will only focus on recommendations for the Commission expert group. The existing framework established by the 2016 Commission Decision is a well thought out structure for the creation and operation of expert groups.[89] The rules provide clarity on how to set up an expert group, explain how it operates and establish important transparency requirements. Informed by this generic framework, the debates on platform councils[90], and the interviews conducted, this section will provide recommendations on how to structure an effective and valuable expert group.

**28. Overview.** After discussing the importance of the selection process (1), this section will suggest the types of experts that should make it up (2).

### a. Selection process

**29. Transparency.** Transparency is a key element to establish trust in the expert group.[91] It is particularly important during the selection process.[92] Accordingly, selection and exclusion criteria should be openly and clearly defined in advance and, ideally, discussed with civil society. This is essential not only for good information but also to achieve widespread consensus. The selection procedure should be as transparent as possible and include, for instance, the total number of applicants.[93] Rejected applicants should be informed of why their application was unsuccessful.

**30. Broad communication strategy.** To make the group as representative as possible, two factors recurred frequently during the interviews: communication and time. First, a good communication strategy can help the call reach members less known to the classical European sphere.[94] Relaying the call via umbrella groups, such as BEUC or EDRi, or via Digital Services Coordinators can contribute to this strategy. Second, providing a broad timeframe will help the call for experts reach different spheres. Organisations also need time to prepare their applications.[95] Under the Commission framework, at least four weeks should be allowed between the publication of the call and its closure.[96] This deadline should be considered a strict minimum and the Commission might want to consider extending the call to at least six weeks.

**31. Continuous open call.** If the first call for experts does not bring a good representation of interests covered by the DSA, the Commission should consider leaving the call open and let new organisations apply on a regular basis.[97] In addition, it might be relevant to re-open the calls for application periodically to allow unrepresented societal interests organisations to apply.

## b. Types of experts

**32. Overview.** The selection of committee members is one of the key elements ensuring the value of the group. It is indeed crucial to guarantee the representativeness of all interests covered by the DSA's topics. This will provide legitimacy to the group but also make its work more useful for all parties involved.

**33. Civil society organisations.** A broad range of civil society organisations should be represented in the expert group. Various organisations have direct experience with topics covered by the DSA.[98] Those with a legal and policy focus can shed new light on some discussions and the monitoring of platforms. Community driven organisations including experiences groups advocating for human rights, gender equality, and non discrimination can provide relevant perspectives and evidence about the effects of technology and online services on these communities.[99] Organisations with a technical background and expertise can provide precious analysis and bring technical evidence of eventual wrongdoing. Also, organisations working on trust and safety, such as former content moderators or integrity professionals, can bring their own expertise.[100] Moderators' trade unions can also function as a "reality check", bringing in a perspective that is often not captured by other stakeholders. Therefore, the call for applicants should be as broad as possible and allow a diverse range of civil society organisations to contribute to the discussions and actions. Also, the selection criteria should be

---

designed to bring in different voices,[101] as well as target various areas of expertise and diverse backgrounds.[102] Ideally, the selection criteria should be established with the involvement of civil society.

A good balance of interest's representation will be important in this group to provide a good balance of interests as well as a diversity of views.

**34. Independent experts and academics.** Individual experts, such as independent experts and academics, can also bring a relevant perspective and a valuable diversity of views.[103] Therefore, individual experts should also be able to apply, provided they are independent and have expertise in fields relating to the DSA.

**35. Industry limited representation.** Here, the notion of industry should be considered broadly, including, for instance, platforms, media companies, advertisers, and their representatives.[104] There are multiple reasons why industry should not have permanent representation in this expert group.

First, industry, namely intermediary services, is subject to the regulation. Because the expert group's missions related to enforcement against this industry, it inherently precludes their representation in the group. As pointed out during multiple interviews, an industry presence can bring an interesting perspective to the work of expert groups when discussing policy making and focusing on specific compliance foundations such as the definition of standards, adoption of codes, etc.[105] There, industry brings significant and useful inputs in the framing and editing of documents. However, during the enforcement process, notably when deciding which practices or actors should be focused on and investigated, industry should not be able to contribute to the discussions. Other reasons include the financial interests of the industry (inherently incompatible with enforcement initiatives),[106] and competition issues such as the implications of deciding if the regulator will monitor potential competitors.[107]

Second, industry is well represented in European and national instances, notably in committees and expert groups,[108] and dedicates a considerable amount of its budgets to gathering information and making connections with policymakers to promote its arguments.[109] The negotiation of the DSA was no exception and Big Tech was heavily involved in lobbying it.[110] Letting industry be represented in this expert group creates the risk of corporate capture, which can have adverse consequences.[111]

Third, multiple interviews reported that industry representatives sometimes do not contribute to the expert process and discussions in a genuine way.[112] For instance, in some expert groups, industry is influencing the priorities of the European Commission by being able to contribute to the drafting of the agenda of key events and by being overrepresented during the vote on the year's priorities.[113] In one specific expert group, the industry representatives came to a meeting and listened to the discussions but did not agree to answer questions from the institutions or other experts, reducing the meeting's value.[114]

Finally, the aim of this group might be put in jeopardy by a body that is overpopulated by stakeholders from various and conflicting interests.

At the same time, the complete exclusion of industry can be counterproductive because, as previously mentioned, it can bring an interesting perspective. Therefore the group should be encouraged to invite, when necessary, industry representatives as external experts.[115]

**36. No need to involve representatives of Member States.** With the creation of the Board for Digital Services,[116] the DSA already provides a framework where Member States and the Commission have a forum to cooperate.[117] Therefore, and for the sake of keeping the size of the group reasonable, representatives from Member States should not be involved in the EU expert group.

---

108  Sarah Arras and Jan Beyers, "Access to European Union Agencies: Usual Suspects or Balanced Interest Representation in Open and Closed Consultations?", *Journal of Common Market Studies*, 2019, vol. 58, no 4, p. 837.

109  Many studies have focus on the lobbying at the European Union legislative level, see for instance, Christophe Crombez, "Information, lobbying and the Legislative Process in the European Union", *European Union Politics*, 2002, vol. 3, p. 7 s.; Heike Klüver, Cælesta Braun and Jan Beyers, "Legislative lobbying in context: towards a conceptual framework of interest group lobbying in the European Union", *Journal of European Public Policy*, 2015, vol. 22, p. 447.

110  Clothilde Goujard, "Big Tech accused of shady lobbying in EU Parliament", Politico, 14 October 2022; Margarida Silva, Luisa Izuzquiza and Felix Duffy, "The Digital Services Act. A case–study in keeping public in dark", EU Observer, 5 July 2022.

111  Interview with Estelle Massé and Eliska Pirkova, 12 January 2022.

112  Interview with Asha Allen, 18 November 2022; interview with Lucie Audibert, 21 November 2022.

113  Interview with Mélissa Chevillard, 18 November 2022.

114  Interview with Lucie Audibert, 21 November 2022.

115  Interview with Tanya O'Carroll, 23 November 2022.

116  Under article 62 of the DSA, the Board is composed of Digital Services Coordinators.

117  Art. 61 of the DSA.

**37. Board.** Under the DSA, the Board will have to significantly involve civil society organisations in its activities.[118] The Board is indeed strongly encouraged to "cooperate with (...) advisory groups".[119] The advisory group could be a way for the Board to develop its cooperation strategy and have an overview of enforcement developments at the European level.[120] Having at least one representative of the Board in the expert group can help develop a coherent and well-designed enforcement strategy. It also avoids scattering ideas and energy. Therefore, the Board's representatives should be invited to contribute to the expert group.

**38. Commission.** Since the main objective of this group is to contribute to the enforcement strategy and particularly the Commission's enforcement work, the Commission must be represented. Traditionally, "expert groups provide advice and expertise to the Commission and its department".[121] Aside from representatives of the Directorate in charge of the DSA's supervisory and enforcement responsibilities, experts from the European Centre for Algorithmic Transparency should also be able to attend the meetings. This Centre will provide in house technical expertise and centralise research from different disciplines that will certainly contribute to the Commission's supervisory and enforcement powers.[122]

**39. Recommendations of categories of members under the Commission framework.** Under the 2016 Commission Decision, expert groups may be composed of five different types of members (type A, B, C, D, E members), each representing a specific type of group and interest.[123] Following our previous recommendations, the DSA expert group should consist of three types of members.

First, it should be open to "Type A members", namely experts acting independently and in the public interest, including individual academics and researchers.

Second, it should be open to some "Type C members", namely organisations in the broad sense of the word, including associations and non-governmental organisations. However, companies, law firms, and consultancies should be excluded because they are either direct or can be indirect representatives of the industry. Trade unions, especially those representing the interests of moderators, should be encouraged to apply.[124] As for universities and research institutes, they should be allowed to apply but special

---

considerations, such as independence and conflict of interests,[125] should be addressed during the selection process.

Third, it should be open to "Type E members", namely other public entities,[126] allowing the Board to be part of the advisory body.[127]

Ideally, the group should not be over-populated and the total number of experts composing the group should be around 30. It is envisioned that most experts of the group will be CSOs' representatives because they represent a broader group of interests and because they have experience in representing users' interests. A fix-term mandate can be a way to renew the group's composition regularly.

> ### Recommendations
>
> - The selection process should be open, transparent, and based on objective criteria
>
> - The Commission should organise a broad communication strategy circulating the call for experts widely and giving organisations at least six weeks to answer the call
>
> - The expert group should be of a reasonable size (around 30)
>
> - The expert group should consist of independent experts, civil society organisations, the Board, and the Commission. Industry and any indirect representatives should not be represented in the group

---

125  Conflict of interests might materialise in case of industry's financing of research relating to content moderation, artificial intelligence, and related topics.

126  Type E members include a broad variety of "other public entities, such as third countries' authorities, including candidate countries' authorities, Union bodies, offices or agencies and international organisations", see. **Art. 7 § 2 (e)** of the Commission Decision establishing horizontal rules on the creation and operation of Commission expert groups. C(2016) 3301, 30 May 2016.

127  There is no clear indication in the DSA regarding the administrative status of the Board. Some experts have considered that it might have a similar status as the European Data Protection Board and qualify as a "Union Body", see Hannah Ruschemeier, "**Re-subjecting state-like actors to the State. Potential for improvement in the Digital Services Act**", in *To break up or regulate Big Tech? Avenues to constrain private power in the DSA/DMA package*, Max Planck Institute for Innovation and Competition Research and Verfassungsblog, Research Paper no 21-25, p. 53.

## 3. Mandate of the expert group

**40. Role.** Under the Commission's framework, expert groups can serve four potential roles.[128] The expert group for the support of the DSA's enforcement could focus on two of those four purposes. First, the group can have a significant role in supporting the preparation of implementing and delegated acts. These acts are fundamental to the good implementation of the DSA and will therefore be essential for effective enforcement. Second, the group can bring significant assistance to the application of the law itself by contributing to the monitoring and investigation of potential violations, as well as monitoring of broader trends in the ways in which digital services are impacting our society.

**41. Missions.** The missions of the expert group have a decisive impact on shaping the group's work.[129] The missions should be largely defined based on the DSA's provisions, with a residual clause.[130] This residual clause will allow the group to provide *ad hoc* opinions on issues relevant to the DSA's enforcement and helps anticipate any potential need relating to the application of the regulation. Keeping the missions of the expert group broad provides more flexibility, especially at the beginning of the mandate. Also, this allows the expert group to agree, for each term, on its own agenda. To establish as soon as possible a culture of cooperation, the missions should be co-decided with members of the expert groups. We anticipate that in the next few months the topics covered by the Commission will focus on data access, algorithmic auditing, and risk assessment.[131]

**42. Keep the group general and do not establish subgroups.** Since the DSA does not target a specific interest but covers a wide range of topics, one of the risks would be to try to have an equal representation of every interest inside the expert group.[132] Indeed, topics relating to online services range from freedom of speech, platforms' impact on minorities, women, and children, but also consumer protection, and the right to privacy. This means effective enforcement of the rules will require a variety of expertise. When confronted with this issue, the Commission sometimes sets up subgroup to examine specific questions. However, appropriate though this solution might be in other sectors, subgroups for the DSA's enforcement could generate confusion about who oversees what and the residual role of the main group,

---

especially during the first years.[133] Most importantly, the topics covered by the DSA are intertwined and complementary, which demands a global approach.[134] Keeping the group general does not prevent its experts from organising themselves in smaller settings and working on specific missions.[135] Consequently, the group should remain general and be open to cooperation with external expertise when needed.

**43. Cooperation with experts and other expert groups.** The European Commission already has several expert groups dedicated to topics indirectly relating to the DSA.[136] One way to integrate the new expert group into the existing landscape of European expert groups would be to arrange cooperation agreements with them. In this way, the expert group can cover any topic relating to digital services while collaborating with other existing experts' groups on specific topics. The DSA already encourages the Commission and the Board to draw on expertise from existing groups,[137] such as the Observatory on the Online Platform Economy.[138]

Another way to bring specific expertise to the new expert group is to encourage the group to invite experts for specific work.[139]

**44. Opinions of the expert group.** Under the Commission's framework, the primary role of the expert groups is to provide specific advice and expertise to the Commission. In principle, the groups "do not take any binding decisions, but they may formulate opinions or recommendations and submit reports".[140] The fact that the opinions are only informative can generate two problems if the Commission tends to not follow up on recommendations. First, it might create a sense of fruitily among the experts who are devoting time and effort. Second, it might limit the group to being merely a discussion forum. While the Commission has to remain fully independent in its enforcement decisions, it is also important for the efficiency of the group that the Commission explains why a recommendation was not followed.[141] The Commission should therefore put in place feedback loops so the group understands the real impact of its recommendations.

---

133   Otherwise, as pointed out during interviews, the group might be ending up working in silos, interview with Eliska Pirkova, 12 January 2022.

134   For instance, the rules protecting minors from targeted advertising not only refer to the protection of minors online but also better regulation of targeted advertising.

135   As it is already the case in some groups, including for instance the EU Observatory on the Online Platform Economy. Interview with Céline Castets-Renard, 30 November 2022.

136   Expert groups already exist in area such as gender equality (Advisory Committee on equal opportunities for women and men, **E01238**), disability (Disability Platform **E03820**), migrants (Expert Group on the views of migrants in the field of migration, asylum and integration, **E03734**), non-discrimination and diversity (High Level Group on Non-Discrimination, Equality and Diversity, **E03328**), protection of consumers (Consumer Policy Advisory Group, **E03750**). Expert groups specifically dedicated to speech also exist, notably a group on hate speech and hate crime (High Level Group on combating hate speech and hate crime, **E03425**), and one on disinformation (Commission Expert Group on Tackling Disinformation and Promoting Digital Literacy Through Education and Training, **E03781**). Groups working on technology also exist such as on artificial intelligence (High-Level Expert Group on Artificial Intelligence, **E03591**).

137   See **recital 134** of the DSA.

138   See the **EU observatory on the Online Platform Economy**.

139   Under article 15 of the 2016 Commission Decision, the Commission's representative "may invite experts with specific expertise with respect to a subject matter on the agenda to take part in the work of the group".

140   European Commission, Communication to the Commission. Framework for Commission expert groups: horizontal rules and public register, C(2016) 3300, 30 May 2016, p. 3.

---

### Recommendations

- The Commission should charge the expert group to "support the preparation of delegated acts" and the "implementation of Union legislation, programmes, and policies"

- The expert group's mandate should be broadly defined according to the DSA's provisions, with a residual clause

- The expert group's specific missions and priorities should be co-decided among the members of the group

- The expert group should cooperate with existing Commission's expert groups

- The Commission, when it does not follow an expert group's recommendation, should explain the reasons why

---

## 4. Administration of the expert group

**45. The impact of the group depends on thoughtful design.** Participation in expert groups is closely connected to the committee's governance function, policy influence, and status.[142] This relies on thoughtful design,[143] especially of the group's administration. To that end, multiple elements will be discussed, including the secretariat of the group (a), its chairpersonship (b), its logistics (c), operation (d) and budget (e). Also, this section suggests a name for the expert group (f).

---

141   The Commission should explain how the measure it is taking is "finding the best solution in the general interest of the European Union and its Member States", see European Commission, Communication to the Commission. Framework for Commission expert groups: horizontal rules and public register, C(2016) 3300, 30 May 2016, p. 3.

142   Renate Mayntz, "Struktur und Leistung von Beratungsgremien: Ein Beitrag zur Kontingenztheorie der Organisation", *Soziale Welt*, 1977, vol. 28, p. 1 s.; see more recently, Sylvia Veist, Thurid Husted and Tobias Bach, "Dynamics of change in internal policy advisory systems: the hybridization of advisory capacities in Germany", *Policy Sciences*, 2017, vol. 50, p. 85 s.

143   Interview with Nani Jansen Reventlow, 23 November 2022.

### a. Secretariat

**46. Secretariat's role.** Traditionally, secretariats have multiple tasks ranging from "policy entrepreneur's tasks (e.g., setting agendas for the […] meetings, preparing background discussion documents, checking the implementation of the work programme) to facilitatory tasks (e.g., assist and support working groups, advisory groups and the Board), and secretarial tasks (e.g., arranging meetings, seminars, taking minutes, updating the […] website)".[144] Given the variety and importance of the tasks handled,[145] the secretariat has considerable influence on the group's deliverables.[146] Consequently, good practices that guarantee a fair distribution of power should be implemented.

**47. Secretariat's composition.** Article 13 of the Commission Decision leaves the group to decide whether it sets up its own secretariat, while requiring the Commission to provide support in case it is needed.[147] Accordingly, the expert group is free to decide how the secretariat should be composed and organised, offering multiple set-up options. A mixed secretariat composed of the Commission's agents as well as representatives of civil society organisations could be a strategic choice. When the tasks are mainly secretarial and logistical, such as arranging meetings, taking minutes, and updating the website, the Commission's agent could be in charge. When the tasks are more strategic, such as setting agendas, inviting external experts, or preparing background discussion documents, it could be useful to put in place a partnership between one agent from the Commission and one elected member of the expert group.[148] Such a partnership can have a positive impact for the whole group because a better distribution of power makes it more engaging for everyone.

**48. Crowdsourcing and review of documents.** It is recognised that "the one holding the pen has far more influence than most other members of a committee".[149] To compensate for the secretariat's unilateral power and to make sure members of the group have their voice represented, influential documents such as the agenda, background discussion documents, and reports should be designed in a collaborative manner, allowing the free flow of ideas.[150] Minutes and documentation should be submitted for review to the group's members before publishing. Involving members at an early stage of the draft

---

offers them time to provide feedback and foster cooperation based on dialogue and trust.[151]

**49. Timing is key.** Anticipation is a characteristic of good governance and helps avoid work overload. Since experts usually have a limited amount of time and resources, one good practice is to give them an indication of the expected workload. As much as possible, it will be necessary to provide transparency about the amount of work required throughout the year, giving experts the opportunity to plan.[152] Also, meetings should be anticipated and scheduled in advance, especially in person meetings.[153] Documents should be sent beforehand to allow comments and iteration. For instance, the final agenda should be sent at least three weeks in advance.[154] Meeting documents should be sent at least two weeks prior to the meeting. These deadlines should be formalised by group decision.[155] Giving the experts time and opportunity to review documents will only be possible if the Secretariat is well organised, but it is crucial.

> ### Recommendations
>
> ● The secretariat should consist of the Commission's agents as well as representatives of civil society organisations
>
> ● The secretariat should involve experts in the drafting process of the agenda and documentation. Experts should also be able to comment on and amend the minutes of meetings
>
> ● The secretariat should give an indication of the anticipated workload

---

151  Sarah Arras and Jan Beyers, "Access to European Union Agencies: Usual Suspects or Balanced Interest Representation in Open and Closed Consultations?", *Journal of Common Market Studies*, 2019, vol. 58, no 4, p. 837.

152  Almost all experts interviewed for the preparation of this report agreed on the fact that anticipation and reasonable delays were key for the expert group's work.

153  The Secretariat should plan meetings at least six weeks in advance, to give time for organisations representatives based far from Brussels to schedule their travel.

154  Interview with Mélissa Chevillard, 18 November 2022; interview with Lucie Audibert and Tomaso Falchetta, 21 November 2022; interview with Estelle Massé and Eliska Pirkova, 12 January 2022. The Multistakeholder expert group to support the application of the GDPR has imposed a four week communication deadline for some of its documents, see Minutes of the 4th meeting of the Multistakeholder expert group to support the application of the Regulation 2016/679, 5 March 2019.

155  Interview with Eliska Pirkova, 12 January 2022.

### b. Chair

**50. Chair's role.** The chair, as the secretariat, has a key position in managing committee work and in shaping its outcome.[156] Traditionally, their main responsibility is to establish the broadest possible agreement in as few meetings as possible.[157] Under article 12 of the Commission decision, the expert group is chaired by "a representative of the Commission or by a person appointed by the Commission". Thus, in principle, the Commission has a prominent role in deciding who will hold this greater influence in the group. The right to appoint the chairperson gives the Commission a powerful instrument.[158] Article 12 of the Commission decision also establishes the possibility of the group "elect[ing] its chairperson by a simple majority of its members". This is a good practice, helping decentralise the group's power and serving the interest of the whole group.

**51. Joint chairpersonship.** Another good practice could be to have the group co-chaired by two persons: one representative of the Commission or the Board and one representative of civil society organisations.[159] This distribution helps to build better interaction between institutions and civil society organisations. A joint chair also increases trust in the expert group, both from the public and its members. Another good practice could be that the joint chairpersonship operates on a rotating basis.[160]

> **Recommendation**
>
> ● The group should be jointly chaired by one representative of the Commission or the Board and one representative of civil society organisations

### c. Logistics

**52. Overview.** To reach experts from outside Brussels and encourage their participation, as well as to accommodate experts with particular needs, it is important to put in place a series of inclusive measures. In other words, the logistics of this group should be as supportive as possible and built on the needs of its experts.

---

156 Eva Krick and Åse Gornitzka. "Tracing scientization in the EU Commission's expert group system". *Innnovation: the European Journal of Social Science Research*. 2020. p. 8.

157 Torbjörn Larsson. *Precooking in the European Union. The world of expert groups*. Report to ESO the expert group on public finance. 2003. p. 73.

158 Torbjörn Larsson. *Precooking in the European Union. The world of expert groups*. Report to ESO the expert group on public finance. 2003. p. 17 and p. 73.

159 Interview with Tim Hughes, 30 November 2022.

160 Interview with Asha Allen, 18 November 2022.

**53. Format of the meetings.** The COVID-19 pandemic has made virtual meetings much more acceptable and has opened up new opportunities for inclusion.[161] Virtual meetings allow easier and increased participation by reducing the time and cost of travel. In person meetings, on the other hand, represent important networking opportunities and allow an easy flow of information between members.[162] Here again, thoughtful design will be vital to make meetings as fruitful and inclusive as possible.[163] A series of recommendations will be discussed.

First, as various interviews pointed out,[164] the group's efficiency will depend on the level of trust built into it. Building trust is an ongoing process, requiring time and voluntary actions, which can be nourished by good practices.[165] For instance, the first meetings could be organised in person to ease initial interactions, build the relationship, and allow trust to thrive, while later meetings could be organised either remotely or in person, depending on the topic and the necessity.[166] Shorter meetings (for instance of one or two hour) can easily be organised remotely, allowing expert members to contribute efficiently while avoiding travel.[167] When in person meetings are organised, a hybrid option should be offered to representatives who cannot attend in person.[168] The success of hybrid gatherings depends on various elements such as making sure experts following remotely feel included,[169] having meeting rooms with adequate sound systems, and regularly opening the floor for discussions.[170] Here again, anticipation and scheduling will be key to allow effective participation from a broad range of actors. Except for *ad hoc* meetings based on specific recent events, meetings should be scheduled at least six weeks in advance, to provide time for experts to plan their trip and contribute to the preparation of documents.

**54. Frequency of the meetings.** Frequency obviously depends on the group's missions. Many interviewees pointed out that meeting regularly is important to build and develop trust and to enable the right flow of information.[171] However, the meetings' frequency should not be decided arbitrarily but should be based on the need for enforcement and regulatory dialogue. When there is a big enforcement burden, meetings should be more frequent; when it is lighter, meetings should be fewer. More importantly, multiple experts have pointed out that meetings are only one part of what makes a group efficient. What happens in between the meetings is critical because it is what drives positive synergies.[172] Therefore, to facilitate communication between meetings, tools should be put in place to allow a good flow of information. For instance, creating a platform for the sharing of information and documents could be very useful.[173] It will also be convenient for asynchronous conversations between group members.

> ### Recommendations
>
> ● The expert group should hold both in person and remote meetings. In person meeting should be preferred at the early stage but as the group moves forward, remote meetings may be favoured
>
> ● The frequency of meetings will depend on workload and enforcement needs. Outside meetings, experts should be able to communicate and share documents easily

## d. Operation

**55. Consensus.** As pointed out previously, the work carried out by the expert group has to be done collectively and in a spirit of mutual trust. The general principles formulated by the Commission framework are coherent with such principles.[174] According to these principles, the group should by default adopt its opinions, recommendations, or reports by consensus.[175] More precisely, "in the event of a vote, the outcome of the vote shall be decided by simple majority of the members. The members that voted against

---

or abstained shall have the right to have a document summarising the reasons for their position annexed to the opinions, recommendations or reports".[176] While consensus is an important principle reflecting the values of trust and community, it is also important to make sure there is no deadlock in the group's work. To help overcome this, the following recommendations on how to reach a consensus can be helpful.[177]

Under the framework, it is unclear if the Commission's representative has voting rights. As this expert group will mainly be dedicated to supporting the Commission's enforcement missions, it might be relevant to give the representatives of the Commission one vote in total, so the participants can be informed of the Commission's view.

**56. Transparency.** To build public trust in the expert group, their work should follow the framework's decision on transparency.[178] By default, "documents of expert groups and sub-groups, including the agendas, the minutes and the participants' submissions" are made available to the public.[179] Exceptions to this principle are limited and comparable to the rules regulating public access to documents, namely protection of a public or a private interest.[180] For instance, the group does not have to publish a "document where disclosure would undermine the protection of: (…) the purpose of inspections, investigations and audits". Because of the specificities of this expert group, which is precisely to support the enforcement missions of the Commission, difficulties will probably arise for documents relating to monitoring or potential investigation. Experts and the Commission will probably have to publish these documents in a timely manner to avoid disrupting the investigation.[181] Another approach, notably for the agenda, can also be to have a public and non-public part of the document. Drawing the line between staggered disclosure and general disclosure will probably be difficult. Members will have to make joint decisions about the sensitivity of their documents and when to make them available to the public. In any case, by default, documents should be published.

---

### Recommendations

- In principle, the expert group should adopt its decisions by consensus
- By default, the group's work should be transparent, and the expert group's documents should be published and accessible to the public

---

### e. Budget

**57. Overview.** Most of the experts interviewed have emphasised how important the group's organisation, influence, and budget are for the involvement of experts. The budget and financial contribution is especially important for smaller organisations and those based outside Brussels.[182] Indeed, economic aspects have an important impact on the inclusivity and representativeness of the group. They should therefore be thoughtfully designed.

**58. Expenses.** By default, the Commission Decision explicitly addresses the economics aspects of the expert groups by allowing reimbursement for "travel and subsistence expenses incurred by participants".[183] Reimbursement traditionally means that the experts must advance the expenses. By offering reimbursement, the Commission outsources the work of booking accommodation and travel. Experts have pointed out that this is an additional burden for them, and have asked for the Commission to offer to take care of it.[184] Some expert groups adopted this good practice. For instance, when the multistakeholder expert group to support the application of the GDPR was established, the Commission booked the accommodation.[185] For experts working with the European Center for Disease Prevention and Control, travel tickets as well as accommodation were booked directly by the institution. This is a good practice which leverages time for the experts, allowing them to focus on the group's work instead of logisitics.

**59. Compensation.** Article 21 of the Commission Decision states that "in principle, participants in the activities of an expert group [...] shall not be remunerated for the services they offer." Hence, the standard call for experts includes a statement that experts are not paid for their time but are only reimbursed for travel and subsistence expenses.[186] By contrast, some expert groups offer reimbursement for expenses as well as compensation. This compensation can reach 450 euros for a daily allowance (and 225 euros for a half day).[187] Studies have shown that calls for expressions of interest with compensation tend to have a high response rate and allow the institution to be more selective during the selection process.[188] This is the case for the "List of Individual Experts for the Implementation of the EDPB's Support Pool of Experts"[189], the "List of

---

Individual External Experts to Assist ENISA"[190] and the "Funding & Tenders Portal Expert Database".[191] Many experts interviewed have pointed out the importance of receiving compensation for the work done within the expert group. This compensation helps experts dedicate the right amount of time to the group's work and foster inclusiveness in the selection process.[192] Remuneration also helps avoid draining resources from civil society organisations. Some interviews also pointed out that it might be less important for bigger structures where the job already *de facto* includes this type of contribution.[193]

In some existing groups, a puzzling system is in place where experts do not receive compensation for their work but consultants supporting their work are remunerated.[194] This is peculiar because the expert's knowledge and work is not valued at the same level as the consultants', even though it is the expert's knowledge that is translated into the group's reports.

Ideally, experts should be able to be remunerated for their work. However, because the creation of the expert group is not explicitly mandated by the DSA, it might be difficult for the Commission to compensate all of them. Therefore, on a trial basis, it could be useful to authorise organisations and individual experts to ask for a special allowance. Not all experts would have to be remunerated but the ones needing such compensation would be able to ask for it.[195] A similar system was put in place by the Regulation establishing the European Securities and Markets Authorities, where members in their expert groups "representing non-profit organisations or academics should receive adequate compensation in order to allow persons that are neither well-funded nor industry representatives to take part fully in the debate on financial regulation".[196]

**60. Origin of the group's funding.** Having to pay for the expert's expenses and stipends might create a burden on the budget of the Directorate in charge of the DSA. Fortunately, there is a way for the Commission to fund the experts' group. Under article 43 of the DSA, the VLOPs and VLOSEs must pay an annual fee which is designed to cover the

---

Commission's costs incurred in relation to the supervisory tasks,[197] including the designation of VLOPs[198]. Because the main missions of the advisory body will be rooted in the DSA's enforcement, it makes it possible to dedicate some of the supervisory fee's to the financing of the expert group.

> ### Recommendations
>
> ● The Commission should, as in any other group, reimburse experts' expenses. Ideally, the Commission should offer to book travel and accommodation
>
> ● The Commission should offer the experts working in the group the opportunity to ask for compensation
>
> ● The Commission can fund the group with the supervisory fees paid by VLOPs and VLOSEs

## f. Name

**61. Name of the group.** Under article 6 of the Commission Decision, the expert group should include, "as far as possible" the term "expert group". To comply with this requirement and to be as self-descriptive as possible, the expert group could be named the "Expert group to support the effective application and enforcement of the Digital Services Act". In public documents, the Commission has referred to an "Expert Group on Digital Services", which could also be a good name.

> ### Recommendation
>
> ● The experts' group can be named the "Expert group to support the effective application and enforcement of the Digital Services Act"

---

197  Art. 43 of the DSA explicitly refers to article 56 and the section relating to the supervision, investigation, and enforcement of VLOPs and VLOSEs.
198  Art. 43 of the DSA explicitly refers to article 33 relating to the designation of VLOPs and VLOSEs.

# B. Selected other cumulative mechanisms

**62. Overview.** As presented before, there are various ways to involve civil society organisations in the institutions' work. This section will briefly discuss a few good practices applicable to traditional cooperation mechanisms, such as public consultations (1). It will then provide recommendations for effective complaints mechanisms (2) as well as less formal cooperation mechanisms such as crowdsourcing events (3).

## 1. Meaningful public consultations

**63. Meaningful public consultations.** Since public consultations are enshrined in the EU Treaties, they are one of the most active tools used by institutions to involve stakeholders in policy making.[199] The first public consultation on the implementation of the DSA was launched in December 2022 on methodology and procedures to calculate the supervisory fee. Many more will be organised to decide on the extent of data access, algorithmic auditing, guidelines for the codes of conduct, and the elaboration of risk assessments.[200]

Because consultation is a widely used mechanism of cooperation, there are many reports that look at improving the effectiveness of public involvement in consultations. Often, these reports highlight the criteria for productive, long term, trusting relationships.[201] Some of the recommendations include, for instance, providing clarity about the aims of the consultation[202] and the relationship to the larger decision making process.[203] They also often recommend presenting information clearly, honestly, and with integrity;[204] as well as some procedural rules promoting power and information sharing among and between participants and decision makers.[205]

> **Recommendations**
> ● Public consultations should follow good practice, including transparency of the process and disclosure of the impact the contributions had on the decision-making process

## 2. Complaints

**64. Complaints as a powerful informative tool.** Complaints are a great way for individuals and civil society organisations to notify regulators about potential violations. For instance, almost a third of the investigations carried out in 2020 by the French data protection authority (the CNIL) were triggered by a complaint.[206] Complaints also allow the parties to have access to the investigation's files and be part of a procedure.[207] Authorities have to adopt processes to deal with complaints and allegations in an efficient way. As pointed out in an OECD report, "it is essential that agencies filter first between complaints that clearly point to possible violations and those that just express some discontent with a business operator but without indication of a regulation being infringed, then between complaints that appear well substantiated and detailed and those that seem less grounded, between those that point to potentially major risks and others that only relate to relatively minor issues, and finally between repeated complaints from several sources and one-off allegations".[208]

**65. Complaints in the DSA.** Even if the DSA does not recognise as many individual rights as the GDPR, the Regulation still recognises some rights such as the recipients' right to contest decisions taken by the providers.[209] Therefore, the complaint mechanisms are mostly organised between the recipients of the service and the service itself.[210] However, recipients can still lodge a complaint against providers of services with the Digital Services Coordinator of their residence, which can if necessary, inform other competent DSCs.[211] No similar mechanism exists for the Commission, which does not have to be informed by DSCs under the provisions nor seems obligated to receive complaints. This appears to be confirmed by the fact that the right to be heard and to access the file are limited to the VLOPs and VLOSEs.[212] In the absence of an operative complaint mechanism with the Commission, a crucial success factor of the enforcement *vis-à-vis* VLOPs & VLOSEs will be for the DSCs to effectively be in a position (with legal empowerment and adequate resources) to contribute to the monitoring;[213] and that the Commission effectively relies on these inputs.

The rights recognised by DSA can be exercised directly by the recipients or indirectly when mandating an organisation to act as "representative".[214]

**66. Complaints at the national level.** Since national regulators are mandated to receive complaints, they will have to put in place mechanisms to do so. Some commentators have already made recommendations for the measures that should be implemented by these "complaints office". They include adding secure channels for sensitive communications, thus enabling whistleblowers to report grievances anonymously, securely, and easily.[215] Similarly to the UK police super-complaints system allowing designated organisations to raise issues on behalf of the public about harmful patterns,[216] trusted flaggers' complaints should receive special treatment. Indeed, the status of trusted flagger is awarded by DSC when an organisation has "particular expertise and competence for the purposes of detecting, identifying and notifying illegal content"[217] and their experience in dealing with platform and removal (or not) of content will be very valuable for the DSCs. Therefore, they should be able to inform DSCs of potential systemic violations and the DSCs should expedite their complaints.

**67. Complaints at the European level.** Proper enforcement of the DSA requires a good information flow, and the Commission will need to collect information from various actors. Even if the DSA does not seem to compel the Commission to receive complaints, the Commission can still implement a platform similar to the ones set out by DSCs. Indeed, complaints are very helpful for the institutions to flag violations on a granular level. It can also add another layer of information when national DSCs are not cooperating readily with the Commission. The Commission cannot remain remote from the evidence from the ground and should open a "tip-line" or a way to receive information.[218] After a pre-selection of the information received by the Commission's secretary, the expert group can be a good resource to help select the most problematic offences channelled through the line and contribute to the first steps of the investigation.

---

213 Under **article 53** of the DSA, the DSC who receives a complaint may share it with competent national authorities and other DSC but does not seem to have to share it with the Commission.

214 **Recital 149** and **art. 86** of the DSA.

215 Julian Jaursch, "**The DSA draft: ambitious rules, weak enforcement mechanisms. Why a European Platform oversight agency is necessary**", Stiftung Neue Verantwortung, 2021, p. 22.

216 Police super-complaints became operational in 2018 and allow designated entities to submit complaints that receive special treatment. see Gov.uk, **Police super-complaint,** website.

217 **Art. 22 § 2** of the DSA.

218 This "tip-line" should be made as accessible as possible for individuals to share and frame their concerns.

> **Recommendation**
>
> ● National DSCs should put in place efficient mechanisms to receive and process complaints and requests. Complaints from trusted flaggers should be fast tracked
>
> ● The Commission should create a "tip-line" or similar mechanism to receive information and evidence
>
> ● The expert group can help in processing complaints received by the Commission

## 3. Crowdsourcing events

**68. The power of the collective.** With the development of digital, new synergies between institutions and civil society have flourished. Notably, this has taken the form of events involving public authorities and civil society such as hackathons.[219] By working in small groups during an intensive time limited event, parties can analyse issues relating to code or data and generate solutions. Taking place outside formal organisational boundaries, hackathons offer "participatory production and creative, fun work in peer communities, blurring the classical lines between enjoyment and work, freedom and control".[220]

**69. Involving civil society in the enforcement of the DSA with crowdsourced events.** Because digital services touch upon fundamental values, such as freedom of expression and human dignity, content moderation practices are a democratic challenge of global scale. To answer this challenge, every stakeholder must be involved to ensure effective enforcement of the regulation. Naturally, online platforms and regulators are already involved. Recipients of the services, meaning the individuals most affected by the online practices, also have an important role to play. As mentioned, unlike in the GDPR, the DSA does not recognise many recipients' rights, even for enforcement of due diligence obligations.[221] However, as "a data-generating piece of legislation",[222] the DSA could still empower the public to analyse the data published by the various parties concerned with transparency obligations.

**70. Access to reporting information.** Many provisions in the DSA foster a culture of transparency, aimed at both online platforms and authorities.

Providers of intermediary services must publish transparency reports yearly "on any content moderation that they engaged in".[223] Providers of online platforms have additional obligations, such as the number of suspensions imposed on their platforms and the number of disputes submitted to the out-of-court dispute settlement bodies.[224]

Authorities (Digital Services Coordinators, the Board, and the Commission) also have many transparency obligations. For instance, the Board, in cooperation with the Commission, must publish comprehensive yearly reports on risk mitigation.[225] The DSCs shall also draw reports on their activities, including the number of complaints received and an overview of their follow-up.[226]

As these transparency obligations are important to make services and institutions accountable, it is crucial to make sure all transparency provisions are correctly implemented and are delivering useful information to the public. To make the most of the reports, the public should be able to make comparisons between intermediary services. The Commission should take full advantage of its powers to adopt implementing acts and "lay down templates concerning the form, content and other details" of reports published by online platforms.[227] Public consultations and the expert group can contribute shaping these implementing acts.[228]

Crowdsourcing events that involve centralising and analysing data could be organised to help the Commission and Member States' authorities in their implementation and enforcement. Civil society started gathering and crowdsourcing as early as February 2023 and co-created a chart mapping VLOPS and VLOSEs across the EU. Indeed, under the DSA, online services had to publish "by 17 February 2023 [...] in a publicly available section of their online interface, information on the average monthly active recipients of the service in the Union, calculated as an average over the period of the past six months".[229] This chart mapping online platforms across the EU is surely an useful resource for the Commission as a first step in its analysis of the designation of VLOPs and VLOSEs.[230]

---

223  Art. 15 of the DSA.

224  Art. 24 of the DSA.

225  Art. 35 of the DSA.

226  Art. 55 § 1 of the DSA.

227  Art. 24 § 6 of the DSA.

228  As mentioned by Owen Bennett, it is highly possible platforms will not want to be compared to each other and will publish reports that do not facilitate this comparison, which is why it might be important for the Commission to adopt implementing acts that provide guidelines to avoid this situation. Interview with Owen Bennett, 15 November 2022.

229  Art. 24 § 2 of the DSA.

230  Under article 33 § 4 of the DSA, the Commission shall, after having consulted the representatives of Member States, adopt a decision designating VLOPs and VLOSEs.

In subsequent years, similar crowdsourcing events could be organised nationally and at the European Union level to look at the information in the documentation of intermediary services such as transparency reports, terms and conditions, recommender systems options. This could be helpful not only to better understand how services are implementing the due diligence obligations but also to make recommendations for implementing acts to make the transparency reports more valuable for everyone involved.

> ### Recommendation
>
> - The Commission should lay down templates concerning the form, content, and other details of reports published by intermediary services
>
> - Civil society organisations should hold crowdsourcing events to dive into the data made available under the transparency provisions of the DSA

# IV
# List of recommendations

## Creation of expert groups at the national and European level

- The European Commission should establish an expert group

- Member States should establish expert groups

- The Commission should rely on the inputs brought by the national regulatory authorities and expert groups at the national and European level

- The Board should foster a dialogue between expert groups

## Creation of an impactful and valuable expert group with the European Commission

- The selection process should be open, transparent, and based on objective criteria

- The Commission should organise a broad communication strategy circulating the call for experts widely and giving organisations at least six weeks to answer the call

- The expert group should be of a reasonable size (around 30)

- The expert group should consist of independent experts, civil society organisations, the Board, and the Commission. Industry and any indirect representatives should not be represented in the group

- The Commission should charge the expert group to "support the preparation of delegated acts" and the "implementation of Union legislation, programmes, and policies"

- The expert group's mandate should be broadly defined according to the DSA's provisions, with a residual clause

- The expert group's specific missions and priorities should be co-decided among the members of the group

- The expert group should cooperate with existing Commission's expert groups

- The Commission, when it does not follow an expert group's recommendation, should explain the reasons why

- The secretariat should consist of the Commission's agents as well as representatives of civil society organisations

- The secretariat should involve experts in the drafting process of the agenda and documentation. Experts should also be able to comment on and amend the minutes of meetings

- The secretariat should give an indication of the anticipated workload

- The group should be jointly chaired by one representative of the Commission or the Board and one representative of civil society organisations

- The expert group should hold both in person and remote meetings. In person meeting should be preferred at the early stage but as the group moves forward, remote meetings may be favoured

- The frequency of meetings will depend on workload and enforcement needs. Outside meetings, experts should be able to communicate and share documents easily

- In principle, the expert group should adopt its decisions by consensus

- By default, the group's work should be transparent, and the expert group's documents should be published and accessible to the public

- The Commission should, as in any other group, reimburse experts' expenses. Ideally, the Commission should offer to book travel and accommodation

- The Commission should offer the experts working in the group the opportunity to ask for compensation

- The Commission can fund the group with the supervisory fees paid by VLOPs and VLOSEs

- The experts' group can be named the "Expert group to support the effective application and enforcement of the Digital Services Act"

## Other cooperative mechanisms

- Public consultations should follow good practice, including transparency of the process and disclosure of the impact the contributions had on the decision-making process

- National DSCs should put in place efficient mechanisms to receive and process complaints and requests. Complaints from trusted flaggers should be fast tracked

- The Commission should create a "tip-line" or similar mechanism to receive information and evidence

- The expert group can help in processing complaints received by the Commission

- The Commission should lay down templates concerning the form, content, and other details of reports published by intermediary services

- Civil society organisations should hold crowdsourcing events to dive into the data made available under the transparency provisions of the DSA

# ANNEX I
## Overview of the DSA's enforcement system

## A. Overview

**71. Coming from.** Adopted in 2000, the e-commerce directive[231] established harmonised rules to "cover certain legal aspects of electronic commerce in the internal market".[232] The structure of the ECD, the DSA's predecessor, relied strongly on self-regulatory initiatives by covered entities[233] and enforcement by Member States.[234]

**72. Going forward.** The DSA's rules still rely on multiple self-regulatory principles by reinforcing due diligence obligations, transparency reporting, and risk assessments. However, the supervision, investigation, enforcement, and monitoring systems have been substantially revisited. In line with the delicate balance between harmonised and coherent enforcement on one hand and the Member States' desire to exercise power over online actors directly on the other, the DSA reflects an intricate interaction between national enforcement and European authorities. The DSA's enforcement system involves various actors alongside the European Commission in a maze of roles and responsibilities.[235]

---

## B. The general role of Member States in the enforcement

**73. Single point of contact.** To facilitate enforcement, all providers of intermediary services must designate a "single point of contact" for direct communication or, if they do not have an establishment in the Union, designate a legal representative in one of the Member States in which they offer services.[236]

**74. The Digital Services Coordinator.** Each Member State shall designate a Digital Services Coordinator (DSC) who is responsible for matters relating to supervision and enforcement.[237] The DSCs are presented as the main enforcers of the DSA. However, they are largely left outside of the enforcement scheme against VLOPs and VLOSEs. For the supervision of their due diligence obligations, the Commission takes the lead and "has exclusive powers".[238]

**75. Cross-border cooperation.** When an alleged infringement occurs in several member States, the DSA offers multiple cooperation mechanisms for DSCs, including mutual assistance,[239] cross-border cooperation,[240] and joint investigations.[241]

## C. The central role of the Commission against very large providers

**76. Exclusive powers of the Commission.** The DSA expressly provides the European Commission with exclusive powers to oversee the enforcement of the due diligence rules in place for VLOPs and VLOSEs.[242] If the Commission is the DSA's leading enforcer against VLOPs and VLOSEs, it also needs to cooperate with the DSCs and the Board at every step of the enforcement.

---

236  Articles 11 and 12 of the DSA.

237  Art. 49 § 1 allows Member States to designate other competent authorities, notably for specific sectors such as electronic communications, media regulation, or consumer protection (see also recital 109 of the DSA). If they do so, Member States need to designate one of the competent authorities as their Digital Services Coordinator.

238  Art. 56 of the DSA. The Commission must cooperate with the Board and DSCs in various ways, see art. 57 of the DSA.

239  Art. 57 of the DSA.

240  Art. 58 of the DSA.

241  Art. 60 of the DSA.

242  Art. 56 of the DSA.

At the early stage, when the Commission opens a proceeding against a VLOP or VLOSE, it must notify all DSCs and the Board[243] so they can provide any information about the infringement. To do so, the Commission will establish and maintain a secure information sharing system.[244]

During the investigation stage, the Commission will carry out its substantial powers of inspection with the assistance of the DSC in whose territory the inspection is being conducted.[245] The Commission can also appoint experts and auditors from competent national authorities.[246]

Finally, when the Commission adopts a non-compliance decision, it must use the enhanced supervision system relying on the cooperation with the DSCs and the Board.[247] In the same vein, the Commission needs to keep the Board and the DSCs "informed about the implementation of the action plan and its monitoring".[248] If all these measures have proven ineffective, the Commission may ask the relevant DSC to request additional restrictive measures including the temporary restriction of access.[249]

These mechanisms rely heavily on the cooperation of public authorities which risks side-lining other interested parties, experts, or civil society organisations.[250] However, the Commission needs to adopt "implementing acts" defining more precisely some aspects of its enforcement including how it should act during inspections,[251] monitoring,[252] and the VLOP and VLOSE's right to be heard.[253] In these implementing acts, the Commission can find creative ways to cooperate with third parties, including civil society organisations.

## D. The role of civil society organisations in the DSA's enforcement system

### 1. The explicit references of the civil society organisations in the DSA's enforcement

**77. Specific references.** The DSA recognises civil society organisations a potentially important actor in enforcement. Explicit references to civil society organisations are spread throughout the text. In some cases they are specifically targeted:

- in the **compliance phase** where CSOs are mentioned multiple times in reference to platforms conducting risks assessment,[254] drawing up of codes of conduct[255] as well as in drawing up crisis protocols.[256] Multiple CSOs will probably develop an important role as "trusted flaggers".[257] Also, they will be able to play an important role in helping the Commission to develop the Union's expertise and capabilities.[258]

- in the **monitoring phase** where CSOs are mentioned as actors who can have access to specific data from VLOPs and VLOSEs to conduct scientific research.[259]

- in the **enforcement phase** where CSOs are mentioned as potential representatives of services recipients.[260]

However, the actual role of the civil society organisations does not have to be limited to DSA provisions where they are explicitly mentioned. There are multiple other implicit references where CSOs can intervene and offer useful expertise.

---

254 Recital 90 of the DSA.

255 Art. 45 of the DSA which will also have an impact on risk mitigation measures, as well as the reporting framework. Articles 46 and 47 of the DSA respectively relating to codes of conduct for online advertising and accessibility, explicitly cite civil society organisations as actors that can contribute to the drawing up of such codes.

256 Art. 48 § 3 of the DSA.

257 Recital 61 and art. 22 of the DSA.

258 Recital 137 and art. 64 of the DSA.

259 Recital 97 and art. 40 § 4 of the DSA.

260 Recital 149 and art. 86 of the DSA.

## 2. The underlying importance of civil society organisations in the DSA's enforcement

**78. Cooperation.** Public authorities at the national and European levels are strongly encouraged to cooperate with each other.[261] Involvement of interested third parties is also supported at various stages of the implementation of the rules. For instance, when the Commission and the DSCs work on guidelines for the mitigation of specific risks, they must organise public consultations, where civil society organisations will surely play a major role.[262]

**79. Contribution to the DSC's activities.** As noted by some commentors, the DSA is broad in scope and introduces rules closely "related to other areas of law (e.g., data protection, audiovisual media regulation, consumer protection, telecommunications, terrorism content), which are already subject to oversight by independent national regulatory authorities".[263] Civil society organisations will be able to contribute to the national Digital Services Coordinators' enforcement activities. It is important to note that some DSCs will be designated among existing national authorities. We can hope, therefore, that these existing national authorities already have cooperation mechanisms in place to coherently work with CSOs, notably at the enforcement stage. If they do not, the DSA might be a good opportunity for them to develop such cooperation. Nonetheless, the DSA opens room for enhanced cooperation between national authorities and civil society organisations. There are two provisions in the DSA advocating such cooperation.

First, when monitoring the regulation's application, the national authorities[264] can send a request for information to the provider of intermediary services of their competence.[265] These requests can also be addressed to "any other person acting for purposes related to their trade, business, craft or profession that may be reasonably aware of information".[266] The term "any" person is a broad and inclusive reference that can certainly include civil society organisations.

---

261  The cooperation mechanisms between the various actors of the DSA's enforcement are detailed in **article 56** of the DSA.

262  **Art. 35 § 3** of the DSA.

263  B. Zeybek and J. van Hoboken, "**The enforcement aspects of the DSA, and its relation to existing regulatory oversight in the EU**", DSA Observatory, 4 February 2022.

264  **Art. 51** of the DSA. This power is also recognised to the Commission for VLOPs and VLOSEs under **article 67 § 1** of the DSA.

265  **Art. 51 § 1 (a)** of the DSA.

266  **Art. 51 § 1(a)** of the DSA.

Second, the DSA also empowers a "body, organisation or association to exercise the rights conferred by this Regulation" on recipients of intermediary services.[267] Representative actions are an "effective and efficient way of protecting the collective interests"[268] and an important milestone for protecting democratic values. The individual rights recognised by the DSA are limited since they are restricted to the right to lodge a complaint against decisions taken by the providers[269] and an out-of-court dispute settlement.[270] Therefore, the representative powers in the DSA are inherently bounded by the restrictiveness of these rights. Moreover, the obligations resulting from the representative power conferred in the DSA are restricted to the providers' obligation to "take the necessary technical and organisational measures to ensure that complaints submitted by bodies, organisations or associations (…) are processed and decided upon with priority and without undue delay".[271] Therefore, the representative role of CSOs is as limited as the individual rights recognised by the DSA.

**80. Contributions to the Board's activities.** Civil society organisations will also be able to contribute to the Board's activities in multiple ways. Indeed, the Board can "invite experts and observers to attend its meetings"[272] and may also cooperate with "advisory groups, as well as external experts as appropriate".[273] Recital 134 of the DSA strongly recommends the Board to cooperate with "advisory groups with responsibilities in fields such as equality, including gender equality, and non-discrimination, data protection, electronic communications, audiovisual services, detection and investigation of frauds against the Union budget as regards custom duties, or consumer protection, or competition law, as necessary for the performance of its tasks". Furthermore, the Board's tasks explicitly prescribe working in cooperation with relevant stakeholders in developing and implementing "European standards, guidelines, reports, templates and codes of conduct".[274] These stakeholders probably include service providers but also various civil groups such as consumer protection coalitions, freedom of expression advocates, data protection organisation, and non-discrimination groups. Also, the Board might substantially rely on civil society organisations and researchers when identifying emerging issues relating to digital services.[275] Clearly, civil society organisations as well as expert and researchers will be a key part of the Board's activities.

---

**81. The Commission's need to develop expertise.** As it is almost unprecedented for the European Commission to monitor and enforce a regulation against VLOPs and VLOSEs, the Commission currently clearly lacks the expertise to do so in an effective manner.[276] Even Commissioner Thierry Breton explicitly recognised the crucial need for the Commission to develop and increase its technical digital skills in order to enforce the DSA package efficiently. To do so, the Commission must increase its "staffing levels and build up specific expertise".[277] As this will take time, the Commission can also rely on external expertise, including develop cooperation frameworks with researchers, associations, as well as international organisations.[278] The DSA provides a great opportunity to create a fruitful and ambitious dialogue between multi-stakeholders. Various provisions unambiguously refer to these potential cooperation mechanisms. For instance, the DSA requires the Commission, in cooperation with the DSCs and the Board, to "develop the Union expertise and capabilities as regards the supervision of very large online platforms or very large online search engines".[279] To do so, the Commission is encouraged to "draw on the expertise and capabilities of the Observatory on the Online Platform Economy (…) relevant expert bodies, as well as centres of excellence. The Commission may invite experts with specific expertise (…), representatives of Union agencies and bodies, industry representatives, associations representing users or civil society, international organisations, experts from the private sector as well as other stakeholders".[280] The expertise of the CSOs is essential not only for the implementation and enforcement of the DSA but also to foresee developments in the coming years.

---

276  Multiple commentors have pointed out the risks relating to this centralisation of power in the hands of the Commission, see for instance for the DMA, Damien Geradin, "The DMA proposal: where do things stand?", *The Platform Law Blog*, May 2021; see for instance for the DSA, Heleen Janssen and Ben Wagner, "A first impression of regulatory powers in the Digital Services Act", Verfassungsblog, 2021; Suzanne Vergnolle, "Enforcement of the DSA and the DMA. What did we learn from the GDPR?" in Symposium *To Break Up or Regulate Big Tech? Avenues to Constrain Private Power in the DSA/DMA Package*, Max Planck Institute for Innovation and Competition Research and Verfassungsblog, 2021, p. 103 s.; Julian Jaursch, The DSA draft: ambitious rules, weak enforcement mechanisms. Why a European Platform oversight agency is necessary, Stiftung Neue Verantwortung, 2021, p. 6; Ilaria Buri, "A regulator caught between conflicting policy objectives", in Debate *Putting the DSA into practice: enforcement, access to justice, and global implications*, DSA Observatory and Verfassungsblog, 31 October 2022.

277  Thierry Breton, "Sneak peek: how the Commission will enforce the DSA & DMA – Blog of Commissioner Thierry Breton", EU Commission's website, 5 July 2022.

278  In Thierry Breton's word "vetted researchers will gain access to data to conduct research that will support our enforcement tasks", see blogpost, "Sneak peek: how the Commission will enforce the DSA & DMA – Blog of Commissioner Thierry Breton", EU Commission's website, 5 July 2022.

279  Recital 137 and art. 64 of the DSA.

280  Recital 137 and art. 64 of the DSA.

**82. Contributions to the Commission's activities.** There are multiple provisions encouraging cooperation with civil society organisations.

Specifically, the Commission can address a request for information not only to the provider of VLOPs and VLOSEs[281] but also to "any other natural or legal person acting for purposes related to their trade, business, craft or profession that may be reasonably aware of information".[282] The possibility of addressing requests for information to organisations that are technically and legally monitoring platforms offers the opportunity for an informative and potentially fruitful cooperation between the Commission and civil society organisations.[283]

Furthermore, the Commission is also encouraged to interview and take statements from "any natural or legal person who consents to being interviewed for the purpose of collecting information, relating to the subject-matter of an investigation, in relation to the suspected infringement".[284] Another possible involvement could be to designate CSOs as expert, so they can conduct an inspection.[285] This power could, however, have unwanted consequences as civil society organisations are often independent counterpowers and this task may drain their resources or jeopardize their independence. Civil society organisations might be more suitable as independent external experts and auditors to assist the Commission in monitoring effective implementation and compliance as well as to provide specific expertise or knowledge to the Commission.[286] As independent external experts, they can provide evidence-based information ensuring effective enforcement while remaining independent from the investigation itself.[287]

---

**TABLE I:** Explicit references to the involvement of CSOs at various stages of the DSA's implementation

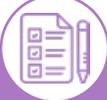

| Compliance stage | Monitoring stage | Enforcement stage |
|---|---|---|
| CSOs can contribute to:<br>● Risk assessments<br>● Drawing up codes of conduct<br>● Drawing up crisis protocols<br>● Developing Commission expertise<br><br>CSOs can be designated as:<br>● Trusted flaggers | CSOs can contribute to:<br>● Conducting scientific research | CSOs can be designated as:<br>● Representatives of service recipients |

**TABLE II:** Implicit references to CSO's involvement at the enforcement stage

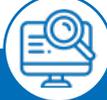

**National authorities**

| | Authorities can cooperate with CSOs by: |
|---|---|
| **Investigation stage** | ● Sending requests for information |
| **Enforcement stage** | ● Receiving observations |

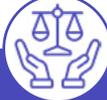

**Board**

| | |
|---|---|
| **Monitoring stage** | ● Inviting experts to attend its meetings<br>● Cooperating with CSOs in its tasks<br>● Developing and implementing standards, guidelines, reports, templates and codes of conduct<br>● Relying on CSOs when identifying emerging issues |

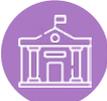

**Commission**

| | |
|---|---|
| **Investigation stage** | ● Sending request for information<br>● Doing interviews and taking statements<br>● Designating CSOs as experts for inspection<br>● Designating CSOs as independent external experts |

# ANNEX II
## List of interviews

## A. Civil society organisations

- **Asha Allen**
  Advocacy Director for Europe, Online Expression & Civic Space,
  **Centre for Democracy and Technology, Europe Office (CDT)**
  Representative at the Advisory Committee on equal opportunities for women and men

  Date of interview: 18 November 2022

- **Mélissa Chevillard**
  Chargée des relations institutionnelles Europe, **UFC–Que Choisir**
  Representing UFC–Que Choisir in the Consumer Policy Advisory

  Date of interview: 18 November 2022

- **Tomaso Falchetta** and **Lucie Audibert**
  Global Policy Lead and Legal Officer, **Privacy International**
  Representing Privacy International in the Multistakeholder expert group to support
  the application of Regulation (EU) 2016/679

  Date of interview: 21 November 2022

- **Nani Jansen Reventlow** and **Jonathan McCully**
  Founder and Head of Legal, **Systemic Justice**

  Date of interview: 23 November 2022

- **Laureline Lemoine**
  Associate, **AWO**

  Date of interview: 4 November 2022

- **Estelle Massé** and **Eliska Pirkova**
  Europe Legislative Manager and Global Data Protection Lead and
  Europe Policy Analyst and Global Freedom of Expression Lead, **Access Now**
  Representing Access Now in the Multistakeholder expert group to support the
  application of Regulation (EU) 2016/679.

  Date of interview: 12 January 2022

- **Diego Naranjo**
  Head of Policy, **European Digital Rights (EDRi)**

  Date of interview: 15 November 2022

- **Agustín Reyna**
  Legal and Economic Affairs Director, **BEUC**

  Representing BEUC in various expert groups

  Date of interview: 9 November 2022

## B. Independent experts

- **Dr Julian Jaursch**
  Project Director, **Stiftung Neue Verantwortung**

  Date of interview: 4 November 2022

- **Tanya O'Carroll**
  Independent advisor

  Date of interview: 23 November 2022

## C. Academics

- **Dr Céline Castets–Renard**
  Professor, **University of Ottawa**

  Member of the EU Observatory on the Online Platform Economy

  Date of interview: 30 November 2022

- **Dr Joris Van Hoboken**
  Professor, **University of Amsterdam**

  Founder of the DSA Observatory
  Former member of the EU Observatory on the Online Platform Economy

  Date of interview: 29 November 2022

## D. Institutions

- **Owen Bennett**
  International Online Safety, **OFCOM**

  Date of interview: 15 November 2022

- **Frédéric Bokobza**
  Deputy Director General, **ARCOM**

  President of subgroup 2, European Regulators Group for Media Services (ERGA)

  Date of interview: 7 December 2022

- **Anna Colaps**
  Member of Cabinet, European Data Protection Supervisor **(EDPS)**

  Date of interview: 22 November 2022

- **European Commission, DG-Connect**

  Date of interview: 31 October 2022

- **Tim Hughes**
  Democracy and Participation Lead, **Open Government Partnership (OGP)**

  Date of interview: 30 November 2022

## Acknowledgements

This report was written by Dr Suzanne Vergnolle with valuable inputs, comments, and analysis from Article 19's team, particularly Chantal Joris and Isa Stasi as well as from Open Society Foundation's team, particularly Guillermo Beltra and Claudio Cesarano. The author is grateful to the many people she interviewed for sharing their experience and providing precious input. She is also grateful to Ilaria Buri, Karolina Iwańska, Soizic Penicaud, and Paddy Leerssen for their useful comments on previous drafts. She also wants to thank Ros Taylor for proof editing the report and Randa Carranza for the design and layout. Finally, she expresses appreciation for funding from Open Society Foundations.

Dr Suzanne Vergnolle

# Putting collective intelligence to the enforcement of the Digital Services Act

While underlying the many ways to build strong cooperation settings between regulators and CSOs, this report focuses on making concrete recommendations for the design of an efficient and influential expert group with the European Commission. The creation of an expert group finds its roots in article 64 and recital 137 of the DSA which require the Commission to develop Union expertise and capabilities. Once established, the experts of this group will be able to bring evidence-based information directly to the Commission and specific expertise on the protection of fundamental rights and the safety of users online. By instituting an expert group, the Commission will not only benefit from valuable expert knowledge but will also demonstrate its willingness to put in place an efficient enforcement system based on collective intelligence.

Aside from the establishment of an expert group, other cumulative mechanisms will also help the DSA's enforcement to thrive. Civil society organisations should, for instance, consider organising regular crowdsourcing events to deep-dive and analyse the data published by entities covered by the transparency obligations. As it has done in the past, the Commission can sponsor these events and be a direct beneficiary of their results. Another way for civil society organisations to bring information to the Regulator is by legal action, including by making complaints to the regulators.